\documentclass{article}
\usepackage{spconf,amsmath,graphicx}

\def\L{{\cal L}}

\title{Learning subject-invariant representations from speech-evoked EEG using variational autoencoders}
\name{Lies Bollens$^{1,2}$, Tom Francart$^2$, Hugo Van Hamme$^1$\thanks{The work is funded by KU Leuven Special Research Fund C24/18/099 (C2 project to Tom Francart and Hugo Van hamme).} \thanks{Research funded by a PhD
grant (1SB1421N) of the Research Foundation Flanders (FWO).} \thanks{This project
has received funding from the European Research Council (ERC) under the
European Union’s Horizon 2020 research and innovation programme (grant
agreement No 637424, ERC starting Grant to Tom Francart).
}}
\address{$^1$KU Leuven, PSI, Dept. of Electrical engineering (ESAT), Leuven, Belgium\\
$^2$KU Leuven, ExpORL, Dept. Neurosciences, Leuven, Belgium
}

\begin{document}
\ninept
\maketitle

\begin{abstract}
The electroencephalogram (EEG) is a powerful method to understand how the brain processes speech.
Linear models have recently been replaced for this purpose with deep neural networks and yield promising results. In related EEG classification fields, it is shown that explicitly modeling subject-invariant features improves generalization of models across subjects and benefits classification accuracy. In this work, we adapt factorized hierarchical variational autoencoders to exploit parallel EEG recordings of the same stimuli. We model EEG into two disentangled latent spaces. Subject accuracy reaches $98.96\%$ and $1.60\%$ on respectively the subject and content latent space, whereas binary content classification experiments reach an accuracy of $51.51\%$ and $62.91\%$ on respectively the subject and content latent space.

\end{abstract}
\section{Introduction}
\label{sec:intro}
Recently, much research has gone into modeling how natural running speech is processed in the human brain.
A standard approach is to present natural running speech to a subject while EEG signals are recorded. Subsequently, linear models are trained to either reconstruct the speech stimuli, to predict EEG from the stimuli, or to transform both stimulus and neural response to a shared space \cite{diliberto_low-frequency_2015, de_cheveigne_decoding_2018, vanthornhout_speech_2018, lesenfants_predicting_2019, di_liberto_neural_2020}. 
Deep neural networks have been proposed as an alternative for these linear models. In \cite{monesi_lstm_2020, accou_predicting_2021, cheveigne_auditory_2021}, the authors relate an acoustic stimulus to EEG using a match/mismatch paradigm and obtain high performance using decision windows of 5 to 10 seconds.

In the related fields of EEG emotion recognition and EEG motor imagery, it has been shown that explicitly modeling subject-invariant features, by using an adversarial layer to remove all subject information from the latent features, improves generalization of models and benefits classification accuracy across subjects \cite{ozdenizci_learning_2020, li_domain_2020, rayatdoost_subject-invariant_2021}. In \cite{li_latent_2020, hagad_learning_2021}, the authors have proposed to include Variational Autoencoders (VAE) \cite{kingma_auto-encoding_2014} and have found that VAE's may improve subject-independent performance, as the latent space is conditioned to follow a Gaussian distribution, and are advantageous in the unsupervised modeling of EEG brain neural signals.  
Though these works yield promising results, little work has been conducted using extremely short windows, necessary for decoding at word-level, as such short windows contain very little information.

Looking at advances in speech processing, a way to accurately model short windows while still keeping track of higher-order features is to use hierarchical models.
A factorized hierarchical variational autoencoder \cite{hsu_unsupervised_2017} (FHVAE) is a deep variational inference-based generative hierarchical model, encoding generating factors into two disentangled latent spaces. One space captures high-level slow varying information, whereas the other captures residual fast-changing information. This hierarchical approach allows capturing high-level information while using small time windows. 

In this work, inspired by recent progress in both speech recognition and related EEG fields, we adopt the architecture of \cite{hsu_unsupervised_2017} and use FHVAE's to model the neural response to a speech stimuli for extremely short segments of EEG of 500~ms.
We propose an extension to the FHVAE architecture, exploiting that the same stimuli can be presented to multiple subjects. 
This work aims to design a model that can generate subject-invariant representations for short EEG segments, as this is a first step towards decoding at the word level. To this end, we measure how well subject and content are represented in their respective latent spaces and how well these two generating factors are disentangled. 
\section{Methodology}
\label{sec:format}

\subsection{Factorized Hierarchical Variational Autoencoders}
\label{sec:fhvae}
\begin{figure}[t]
\begin{minipage}[b]{1.0\linewidth}
  \centering
  \centerline{\includegraphics[width=8.5cm]{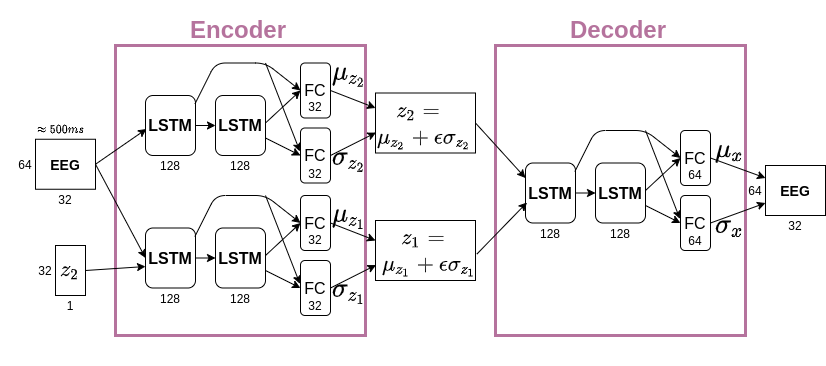}}
\end{minipage}
\caption{Proposed architecture of Factorized Hierarchical Variational Autoencoders. FC refers to a single fully connected layer. We refer to $\textbf{z}_1$ and $\textbf{z}_2$ as the \textit{latent content variable} and the \textit{latent subject variable} respectively.}
\label{fig:fhvae}

\end{figure}

Factorized Hierarchical Variational Autoencoders (FHVAE) \cite{hsu_unsupervised_2017} are a variant of Variational Autoencoders \cite{kingma_auto-encoding_2014} designed to model sequential data using a hierarchical model. The archictecture is shown in figure \ref{fig:fhvae}. Each uninterrupted recorded sequence $i$ of EEG signals $X^i$ can be decomposed into $N$ segments of fixed length $ \{\textbf{x}^{(i,n)}\}_{n=1}^N$. The FHVAE model imposes that the generation of each segment $\textbf{x}$ is conditioned on two latent variables, $\textbf{z}_1$ and $\textbf{z}_2$ (omitting superscripts). We refer to $\textbf{z}_1$ as the \textit{latent content variable}, as $\textbf{z}_1$ is generated from a global prior and captures the fast-changing generating factors between segments. On the other hand, we refer to $\textbf{z}_2$ as the \textit{latent subject variable}, as $\textbf{z}_2$ is drawn from a sequence-dependent Gaussian prior, parameterized by an (estimated) sequence-level mean $\mu_2$ and a fixed variance. 
The FHVAE learns to encode common factors between segments from the same sequence into $\textbf{z}_2$.
By contrast, $\textbf{z}_1$ has a zero-mean prior.

In order to avoid collapse (trivial solution for $\textbf{z}_2$), a discriminative objective is added: it should be possible to infer the sequence number $i$ from $\textbf{z}_2$.
An FHVAE introduces an inference model $q$, since the true posteriors of $\textbf{z}_1$, $\textbf{z}_2$ and $\mu_2$ are intractable. 
More detailed derivations can be found in \cite{hsu_unsupervised_2017}. \\
The objective the FHVAE tries to maximize can be summarized as, with $p$ referring to the prior: 
$$\L^{dis}(p,q;\textbf{x}^{(i,n)}) = \L(p,q;\textbf{x}^{(i,n)}) + \alpha_{z_2} \log p(i \vert \textbf{z}_2^{(i,n)}) $$
where $\L(p,q;\textbf{x}^{(i,n)})$ is the variational lower bound and $\alpha_{z_2}$ a weighing parameter.

\subsection{Extended Factorized Hierarchical Variational Autoencoders}
\label{sec:extfhvae}
It is possible to generate parallel data when recording EEG brain signals. Parallel refers to multiple recordings of one or more subjects listening to the same stimulus. This data organization is not exploited in the original FHVAE. Therefore, we extend the FHVAE by adding an extra regularisation component on the latent $\textbf{z}_1$ space. 
Similar to the approach used to model sequence-level information in the latent $\textbf{z}_2$ space in the original model, $\textbf{z}_1$ is now generated from a content-dependent Gaussian prior with fixed isotropic variance $\sigma_{z_1}^{2}$. 

We replace the prior on $\textbf{z}_1$ by $p(\textbf{z}_1^{(i,n)} ) = \mathcal{N}(\mu_1^{l(i,n)}, \sigma_{z_1}^{2}\pmb{I}) $, while the content-specific means $\mu_1$ are in turn assumed to be distributed as $p(\mu_1) = \mathcal{N}(\pmb{0}, \sigma_{\mu_1}^{2}\pmb{I})$. The other priors stay identical with respect to the original FHVAE. 

Similar to \cite{hsu_scalable_2018}, we introduce a $\mu_1$-table, with one entry for each unique stimulus segment (of 500~ms) $l_{k}$. We will refer to $l_k$ as the \textit{content labels}. As we generate parallel EEG data, multiple EEG segments have the same content label.
For each EEG segment $\textbf{x}^{(i,n)}$, $l(i,n)$ uniquely identifies the associated content label. We define the associated inference model again as Gaussian. 


In the segment variational lower bound, the KL-Divergence for $\textbf{z}_1$ changes into
$D_{KL}(q(\textbf{z}_1^{(i,n)} \vert \textbf{x}^{(i,n)}, \textbf{z}_2^{(i,n)}) \Vert p(\textbf{z}_1^{(i,n)} \vert \mu_1^{l(i,n)}) $ 
We add one extra term to the summation for $\mu_1$:
$\frac{1}{S} \log  p(\mu_1^{l(i,n)})$, where $S$ is equal to the number of occurences of the content label.  
Finally, it is necessary to add an extra discriminative term to the total variational bound in order to avoid collapse for $\textbf{z}_1$. This results in the following final objective, with $\alpha_{z_1}$ and $\alpha_{z_2}$ weighing parameters: 
\begin{align*}
\L^{dis}(p,q;\textbf{x}^{(i,n)}) &= \L(p,q;\textbf{x}^{(i,n)})+ \alpha_{z_1} \log p(l(i,n) \vert \textbf{z}_1^{(i,n)}) \\ 
&+ \alpha_{z_2} \log p(i \vert \textbf{z}_2^{(i,n)}) 
\end{align*}

In order to avoid scalability issues due to a large $\mu_1$-table, we follow the training approach suggested in \cite{hsu_scalable_2018} and implement hierarchical sampling at the content level. Per batch, we choose $K$ unique content labels $\left\lbrace l_k\right\rbrace_{k=1}^{K}$ and include all EEG segments for which the label $l(i,n)$ is part of this set.

\section{Experimental Setup}
\subsection{Data collection and preprocessing}
\label{sec:training}
We present natural running speech to subjects while simultaneously recording the EEG signal. In total, 100 normal-hearing, native Flemish participants contributed to this study. Normal hearing was confirmed using pure tone audiometry and the Flemish MATRIX speech-in-noise test \cite{luts2014development}. We present to each participant a minimum of 6 and a maximum of 8 stories, each around 15 minutes in length, chosen from a set of ten different stories and presented in random order. We present the stories binaurally at 62 dBA with Etymotic ER-3A insert phones and ask the subjects a comprehension question after each story. The EEG data is recorded using a 64-channel Biosemi Active-Two EEG at a sampling rate of 8~kHz. Stories are presented using the APEX 4 software platform \cite{francart_apex_2008}. Subjects are seated in an electromagnetically shielded and soundproof booth. 
We remove artifacts from the EEG recording using a multichannel Wiener filter \cite{somers_generic_2018}. We bandpass filter the EEG signals between 0.5~Hz and 32~Hz and then downsample to 64~Hz. 
We use the first and last $40\%$ for training and divide the remaining $20\%$ equally into validation and test set.

\subsection{Training and Model Configurations}
We present EEG segments of 32 frames, corresponding to 500 ms, as input to the model. We implement the architecture introduced in \cite{hsu_scalable_2018}.
$\textbf{z}_1$ and $\textbf{z}_2$ are predicted by a stacked LSTM network of two layers, each containing 128 cells, followed by two separate single fully connected layers, predicting the conditional mean and variance, respectively. These last layers take as input the output from the last time step from both LSTM layers, summing up to a total of 256 dimensions. 
The decoder network reconstructing $x$ consists of a two-layer stacked LSTM decoder network, which takes as input at each time step the concatenation of the sampled ${\textbf{z}_1}$ and ${\textbf{z}_2}$ from the posterior distribution. Subsequently, the predictor network takes the LSTM output of the last layer from each time step and predicts the probability distribution of the corresponding time frame $p(x_t \vert \textbf{z}_1, \textbf{z}_2)$. 
The dimensions of $\textbf{z}_1$ and $\textbf{z}_2$ are both 32. We set the variances as $\sigma^2_{z1} = \sigma^2_{z2} = 0.25$ and $\sigma_{\mu_1}^2 = \sigma_{\mu_2}^2 = 1$. 

We train the model in two steps. First, we train a default FHVAE, as described in section \ref{sec:fhvae}, allowing the model to focus on modeling subject-specific generating factors in the latent $\textbf{z}_2$ space. We set $\alpha_{z_2}=100$. We will refer to this model as \textit{FHVAE}.
Second, we add the extra regularisation on $\textbf{z}_1$ and train our extended FHVAE model, as described in section \ref{sec:extfhvae}, such that the model can now focus on modeling content-dependent, subject-independent generating factors in the latent $\textbf{z}_1$ space. We start training with the model weights obtained in the first step. We set $\alpha_{z_1}=10000$, $\alpha_{z_2}=100$ and choose a segment batch size $K=5000$. We will refer to this model as \textit{Extended FHVAE}.

We use ADAM \cite{kingma_adam_2017} with $\beta_1=0.95$ and $\beta_2=0.999$ for both training steps and train over 500 epochs, with early stopping when the variational lower bound on the held-out validation set does not improve for 50 epochs. We use Tensorflow 2.0 for implementation. 

\subsection{Subject Classification and Disentanglement} 
\label{sec:subject-experiment}


We describe the setup used to quantify subject classification performance of the model. The question we answer here is two-fold: does the model succeed at extracting the subject generating factors in the $\textbf{z}_2$ latent space and disentangling subject from content information?
we group segments of EEG according to subject, which results in a total of 100 different classes. 
Train, test, and validation sets are taken as the sets used for the training of the FHVAE, described in section \ref{sec:training}. For all segments, we infer the latent $\textbf{z}_1$ or $\textbf{z}_2$ and then train a one-layer fully connected network with softmax activation and 100 output classes, using categorical cross-entropy as the loss function and ADAM as optimizer.

To validate our results, we compare with a recently proposed architecture for EEG person authentication \cite{ozdenizci_adversarial_2019}, which has been compared in a recent survey about methods and challenges in EEG person authentication \cite{jalaly_bidgoly_survey_2020} and yields excellent results. The authors of \cite{ozdenizci_adversarial_2019} propose to use an encoder consisting of a convolutional neural network, followed by a single fully connected layer for subject classification. We will refer to this network as CNN. This architecture uses EEG window lengths of 500~ms, making it directly transferable to our paradigm. We use the preprocessing pipeline of our EEG data and sample at 64~Hz, which differs from the 256~Hz proposed by the authors. We use the same train, test, and validation set as above. We jointly train the CNN encoder and output classification network, using categorical cross-entropy as loss function and ADAM as optimizer.

\subsection{Content Classification and Disentanglement}
\label{sec:content-experiment}
We describe the setup used to quantify content classification performance of the model. The question we answer here is two-fold: does the model succeed at extracting the content generating factors in the $\textbf{z}_1$ latent space and disentangling content from subject information?

We keep all EEG segments of the test set, as described in section \ref{sec:training} and group them according to their content label $l$, which results in a total of 1538 classes. Each class contains, on average, 80 examples. Instead of directly classifying which of these 1538 classes the latent representations belong to, which would be a high-dimensional problem prone to overfitting given the limited training examples per class, we simplify the classification problem into a binary one. 
We select a label $l_k$ at timestep $t_i$ and choose a second label $l_{k+1}$ directly after, at timestep $t_i+500ms$.
We split these two classes into $80\%$ train and $20\%$ test set, using a 5-fold validation scheme. Subsequently, a Linear Support Vector Machine is trained, and we calculate average accuracy over the test set. We conduct this experiment for $k=1..1537$, effectively utilizing the whole test set. 

As a baseline, we take the same approach to selecting classes and training, but instead of inferring the latent variable, we flatten the EEG segments and directly train a Linear Support Vector Machine on the raw EEG. 

\begin{figure}[t]
\begin{minipage}[b]{1.0\linewidth}
  \centering
  \centerline{\includegraphics[width=8.5cm]{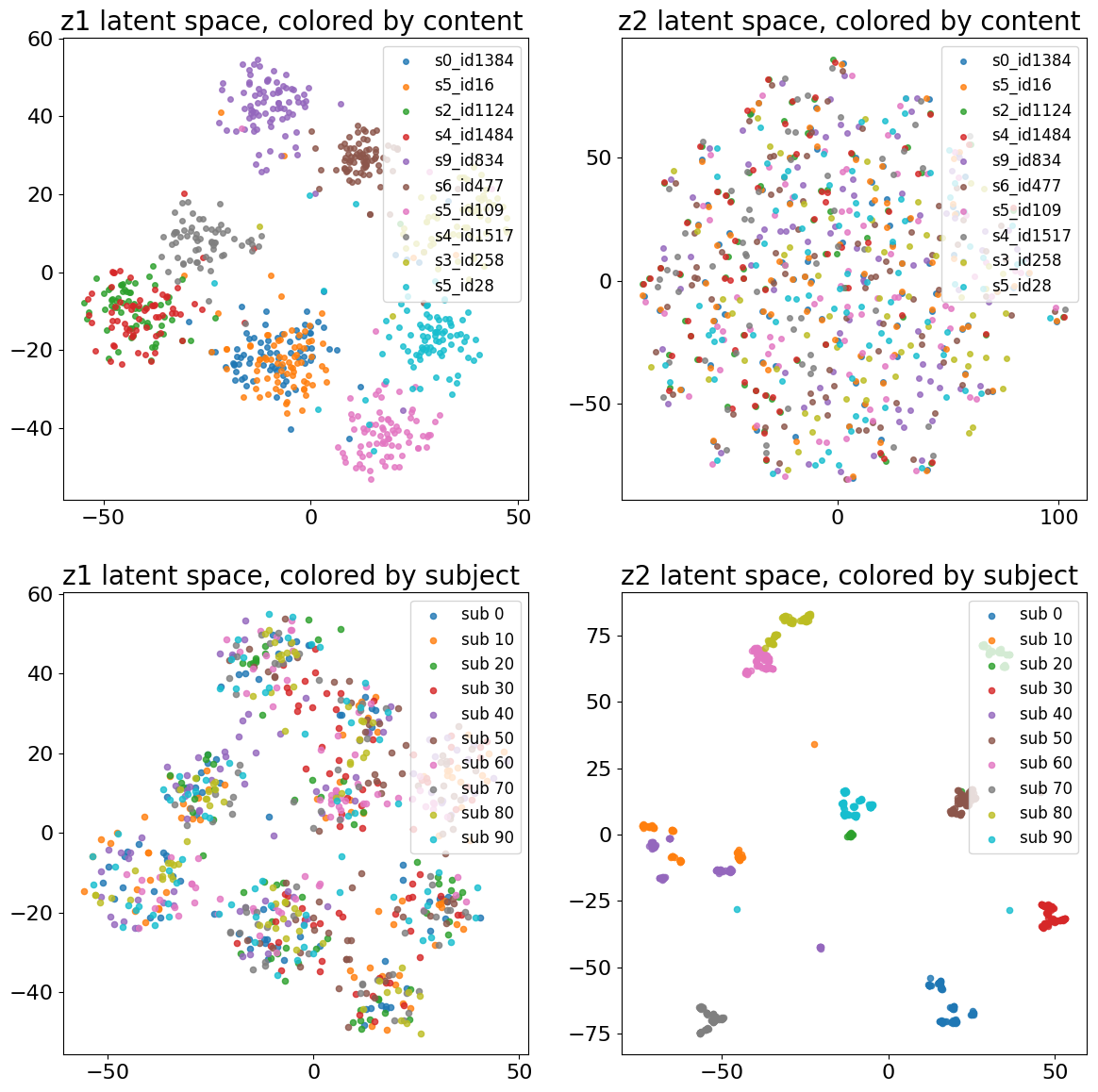}}
\end{minipage}
\caption{t-SNE scatter plots of $\textbf{z}_1$ and $\textbf{z}_2$ latent space. Every point represents one segment. Different colors indicate either content segments or subjects. Different generating factors (subject, content) are shown at the title of each plot}
\label{fig:res}
\end{figure}

\section{Results}

\subsection{t-SNE Visualization of Latent Variables}
\label{sec:tsne}
In this section, we report a qualitative visual inspection to evaluate performance of the proposed Extended FHVAE model containing 128 LSTM cells.
Therefore, we visualize the latent $\textbf{z}_1$ and $\textbf{z}_2$ space according to different generating factors, which can be either subject or content.

We start by randomly selecting 100 content labels and keeping all segments from all of the 100 different subjects that belong to one of these content labels, resulting in a set of 7453 unique labeled segments. We then infer $\textbf{z}_1$ and $\textbf{z}_2$ for each of these labeled segments and project them onto a two-dimensional space using t-Distributed Stochastic Neighbor Embedding (t-SNE) \cite{maaten_visualizing_2008}. For visualization, we color-code the projected $\textbf{z}_1$ and $\textbf{z}_2$ according to their generating factors. A successful model should result in segments of the same subjects forming clusters in the projected $\textbf{z}_2$ space but not in the projected $\textbf{z}_1$ space. On the other hand, segments of the same content should form clusters in the projected $\textbf{z}_1$ space but not in the projected $\textbf{z}_2$ space.

Results of the t-SNE projections are shown in Figure \ref{fig:res}. For clarity, we only visualize segments belonging to a subset of 10 out of the 100 classes. Each point represents one segment. In the plots grouped by subject, the projected $\textbf{z}_2$ space forms visible clusters (bottom right), whereas the projected $\textbf{z}_1$ space (bottom left) shows much less pronounced clusters and seems to have a more even distribution among all generating factor values. This implies that $\textbf{z}_1$ contains much less subject information and disentanglement on the subject level is successful. 
In the plots grouped by content, the opposite can be observed. Segments cluster in the projected $\textbf{z}_1$ space but not in the projected $\textbf{z}_2$ space, implying that no content information exists in the $\textbf{z}_2$ latent space.

\subsection{Subject Classification and Disentanglement: Experiments}
\label{sec:subject-classification}

\begin{figure*}[htb!]
\begin{minipage}[b]{1.0\linewidth}
  \centering
  \centerline{\includegraphics[width=\textwidth ]{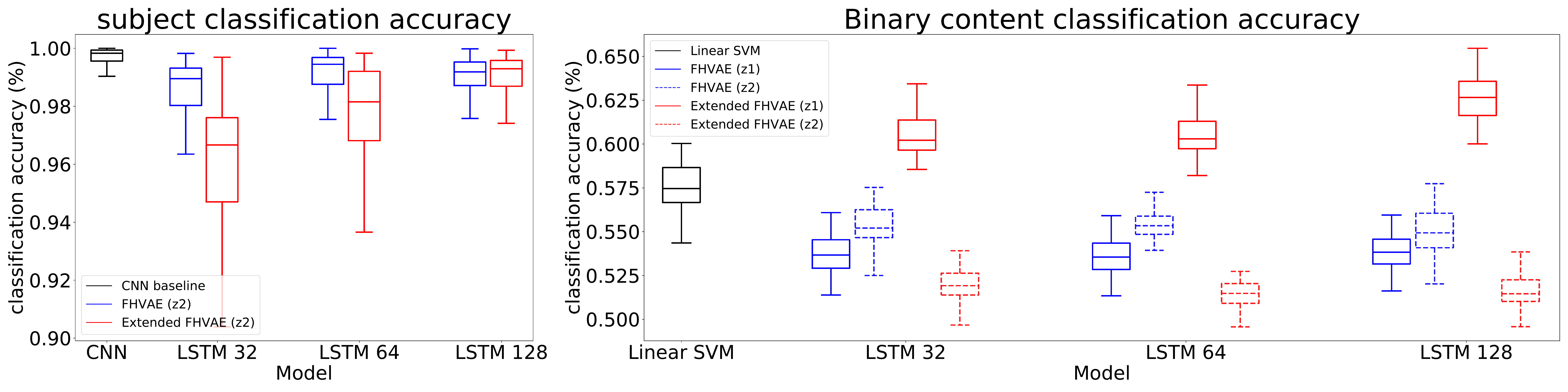}}
\end{minipage}

\caption{\textit{Left}: subject classification accuracy, calculated over all subjects ($N=100$). We compare a state-of-the art CNN model with the default FHVAE and the proposed Extended FHVAE. Results are shown for the latent $\textbf{z}_2$ space for varying model LSTM size. \textit{Right}: Binary content classification accuracy, calculated over all different stories over different folds ($N=45$). We compare a simple Linear SVM with the default FHVAE and the proposed Extended FHVAE. Results are shown for the latent $\textbf{z}_1$ and $\textbf{z}_2$ space  for varying model LSTM size. }
\label{fig:content}
\end{figure*}

In this section, we report subject classification results, as described in section \ref{sec:subject-experiment}. Results are shown in figure \ref{fig:content}. We report the mean accuracy over 100 subjects. For clarity, the subject classification accuracy of $\textbf{z}_1$ is left out of the plot. This accuracy ranges between $1.60\%$ and $2.53\%$ for different models, where $1\%$ is the chance level since all 100 subjects are equally present in the data set. 
During the first stage, while using the original FHVAE architecture, the model should focus on extracting subject information from the EEG segments and disentangling this information from content. Subject classification accuracy on the latent $\textbf{z}_2$ space reaches up to $99.04\%$ for the model using 64 LSTM cells, confirming that during the first stage, the model learns to model subject-specific information in the latent $\textbf{z}_2$ but not in $\textbf{z}_1$, achieving great disentanglement. 

During the second stage, while training the Extended FHVAE architecture, the model focuses on extracting content representations in the $\textbf{z}_1$ space, which is reflected in the subject classification accuracy: for the LSTM model containing 32 or 64 cells, there is a significant ($W=204$, $p<0.001$ and $W=199$, $p<0.001$) decrease in accuracy of $1.46\%$ and $3.23\%$ respectively. We report all statistics using a Wilcoxon-signed rank test with normal approximation. This accuracy drop might be indicative of an inherent trade-off in the model between modeling subject and content.
However, for the 128 LSTM cells model, the mean accuracy of the FHVAE is $98.94\%$, which does not decrease in the second step, from which we conclude that, with regard to subject classification accuracy, there is an advantage to using the bigger 128 LSTM cells model. 
 
The CNN model achieves $99.33\%$ subject classification accuracy, which is slightly yet significantly ($W=750$, $p<0.001$) higher than the extended FHVAE model with 128 LSTM cells at $98.96\%$.

\subsection{Content Classification and Disentanglement: Experiments}
\label{sec:content-classification}

In this section, we report content classification results, as described in section \ref{sec:content-experiment}. Results are shown in figure \ref{fig:content}. We report the average accuracy over stories, over folds. In total, there are nine different stories to which the subjects listen. For the standard FHVAE model, accuracy is about the same for the 32, 64 and 128 LSTM model for $\textbf{z}_1$ ($53.65\%, 53.55\%$, and $53.87\%$) and $\textbf{z}_2$ ($55.31\%$, $55.27\%$, and $55.03\%$). The best performing model of these is still significantly lower ($W=3$, $p<0.001$) than the classification accuracy of the naive SVM ($57.45\%$), from which we can conclude that the standard model achieves neither content extraction nor disentanglement.

However, after training the extended FHVAE model, classification accuracy of the latent $\textbf{z}_1$ space significantly improves ($W=0$, $p<0.001$) from $54.10\%$ to $62.91 \%$ for the model with 128 cells, which indicates that the model succeeds at extracting some relevant content information in the $\textbf{z}_1$ space. The classification accuracy of the smaller LSTM models sees a smaller increase to around $60\%$, from which we can conclude that, with regard to content classification accuracy, there is an advantage to using the bigger 128 LSTM cells model.
For the latent $\textbf{z}_2$ space, performance drops to just barely above the chance level at $51.51\%$, suggesting better disentanglement for content in the extended FHVAE model.

\section{Conclusion}
This work aims to design a model that can generate subject-invariant representations for short EEG segments of around 500 ms, a necessary temporal resolution for decoding at the word level. To this end, we introduce factorized hierarchical variational autoencoders (FHVAE). FHVAE's encode the generating factors of the EEG into two disentangled latent spaces. The latent $\textbf{z}_2$ space should encode subject but not content, whereas the latent $\textbf{z}_1$ space should encode content but not subject. In order to exploit that it is possible to generate multiple EEG recordings of one or more subjects listening to the same stimulus, we propose an adaption to the original FHVAE architecture: Extended factorized hierarchical variational autoencoders. 

We train both models using EEG recordings from 100 subjects who listen on average to 8 stimuli, each around 15 minutes. In a first step, we train the default FHVAE architecture, allowing the model to focus on modeling subject-specific generating factors in the latent $\textbf{z}_2$ space. In a second step, we use the learned model weights and train the Extended FHVAE architecture, allowing the model to focus on modeling content-specific generating factors in the latent $\textbf{z}_1$ space. 

Subject classification accuracy between all 100 subjects reaches $98.94\%$ and $98.96\%$  for the latent $\textbf{z}_2$ representations of respectively the FHVAE and the Extended FHVAE architecture, showing that the default FHVAE model succeeds at extracting subject information and that this information is kept in the second stage. Subject classification accuracy for the latent $\textbf{z}_1$ is for both models at around $2\%$, suggesting great disentanglement with regard to subject. 
At $99.3\%$, subject classification for the state-of-the-art CNN model is slightly superior, but its sole goal is to represent subject, thereby discarding content information. The Extended FHVAE however provides disentangled representations of both content and subject.

Binary content classification accuracy for the default FHVAE model reaches $53.87\%$ and $55.27\%$ for respectively the latent $\textbf{z}_1$ and $\textbf{z}_2$ representations. 
Content classification accuracy for the Extended FHVAE model improves to $62.91\%$ and $51.51\%$ for respectively the latent $\textbf{z}_1$ and $\textbf{z}_2$ representations, confirming that the Extended model succeeds at modeling content generating factors in $\textbf{z}_1$ but not in $\textbf{z}_2$, something which the default FHVAE fails to achieve.

We conclude that the Extended FHVAE model successfully models both subject and content generating factors and disentangles these two while using small EEG segments of around 500~ms. Discrimination between such small-scale segments might be a first step towards decoding at the word level.




\vfill\pagebreak

\bibliographystyle{IEEEbib}
\bibliography{refs}

\end{document}